\documentclass[a4paper,english]{article}
\usepackage[T1]{fontenc}
\usepackage[latin9]{inputenc}
\makeatletter

\usepackage{babel}
\makeatother

\begin{document}

\title{A Simple Proof for the Theorem of Wigner }

\author{Manfred Buth\date{}}

\maketitle
\begin{abstract}
The leading idea of the paper is to treat the theorem of Wigner with
geometrical means and especially to reduce the general case to simple
geometry in two or three dimensions. Thus the special case of three
dimensions is on the one hand the core of the general proof and on
the other hand a special example well suited to illustrate the essential
features of the theorem of Wigner.
\end{abstract}

\section{Introduction}

The theorem of Wigner {[}1] is an important result of quantum mechanics,
and this is its message: If one describes a physical system by the
states of a Hilbert space and afterwards changes to another representation,
then the invariance of transition probability is not only necessary
for the correspondence with the experimental results but also sufficient
for the existence of a transformation of the Hilbert space, which
is either linear and unitary or antilinear and antiunitary.

The aim of the present paper is to present a simple proof of this
theorem with the leading idea to reduce the investigation of the general
case dealing with arbitrary Hilbert spaces to some elementary exercises
in the geometry of two or three dimensions.

In the next section some concepts are introduced in order to facilitate
the formulation of the theorem which is given in section 3. The core
of the paper is the proof of the theorem. It is presented in section
4 and subdivided into six parts. Section 5 contains the comparison
with two other proofs that can be found in the literature, before
finally the essential features of the proof are summarized.

\section{Conceptual preliminaries}

Rays in a Hilbert space are nothing else but one dimensional subspaces.
A ray plane shall be the space orthogonal to a ray, and a ray mapping
be a mapping from the set of all rays of a Hilbert space into the
corresponding set of all rays of another Hilbert space. In order to
define the ray function $u$ for two rays $r$ and $s$ first of all
the expression\[
u(e,f):=<e,f><f,e>/<e,e><f,f>\]
is defined for two arbitraryly chosen elements $e$ and $f$ generating
the two rays $r$ and $s$. Since $u(e,f)$ is independent of the
choice of $e$ and $f$ the ray function $u$ is defined for the rays
$r$ and $s$ by setting\[
u(r,s):=u(e,f)\]
Orthogonality between two rays $r$ and $s$ is defined by the condition
\[
u(r,s)=0.\]

\section{Assumptions and assertions of the theorem}

The theorem of Wigner is presented in the following version:

Let be given two Hilbert spaces $H$ and $H'$ over the field of complex
numbers and an invertible ray mapping $\sigma$ from $H$ onto $H'$
that, together with its inverse mapping, conserves orthogonality.
Then the following assertions are valid: 

(a) $H$ and $H'$ are isomorphic. 

(b) There is a mapping $\varphi$ from $H$ to $H'$ that can be described
by the relations\[
x_{i}'=r_{i}f(x_{i})\qquad\qquad i\epsilon I\]
between the coordinates $x_{i}$ of an element $x$ of $H$ and the
coordinates $x_{i}'$ of the image $x'$ of $x$ under $\varphi$,
provided suitable bases of $H$ and $H'$ with a common set $I$ of
indices are chosen. The factor $r_{i}$ is a positive real number
and $f$ an automorphism on the field of complex numbers.

(c) If additionally the invariance if the ray function is assumed,
then \[
x_{i}'=f(x_{i})\qquad\qquad i\epsilon I\]

and $f$ is either the identity or the transition to the complex conjugate.

(d) The ray mapping $\sigma$ is reproduced by the mapping $\varphi$.

\section{The Proof of the theorem}

\subsection*{Part 1}

First of all the assumption (a) must be proved.

If an orthonormal base $B$ of $H$ over the set $I$ of indices is
given, then the elements $e_{i}$ of $B$ generate rays $k_{i}$ playing
the role of coordinate axes in the sequel of the proof. They are mutually
orthogonal, and hence their images $k_{i}'$ in $H'$ under $\sigma$
are orthogonal, too. Assumed there were a further ray $k'$ in $H'$
orthogonal to all rays $k_{i}'$, then the original $k$ of $k'$
under $\sigma$ would be orthogonal to all rays $k_{i}$ in $H$ and
hence would contain a normed element $e$ being orthogonal to all
$e_{i}$ of $B$. But this is impossible, because the base $B$ was
assumed to be complete. Selecting one normed element $e_{i}^{*}$
out of each ray $k_{i}'$ will give a base $B^{*}$ of $H'$ over
the same set of indices as for $B$. Hence the two Hilbert spaces
are isomorphic.

\subsection*{Part 2}

The main task of the proof is the construction of a mapping $\varphi$
from $H$ to $H'$ which satisfies the assertion (b) of the theorem.

The first step to do this consists in the decree that the element\[
e_{i}'=e_{i}^{*}p_{i}\qquad\qquad i\epsilon I\]
with a phase factor $p_{i}$ shall be the image of $e_{i}$ under
$\varphi$. The factor $p_{i}$ is held open, until it will be fixed
in part 4 of the proof.

Next an index $1$ is selected in $I$ and with it the subset $E$
containing all elements of $H$ with $x_{1}=1$. In $H'$ the subset
$E'$ shall be given by the equation $x_{1}'=1$. A mapping $\varphi_{0}$
from $E$ to $E'$ is defined by the following construction: Any element
$x$ of $E$ is contained in a ray $s$. This ray cannot be orthogonal
to $e_{1}$. Hence the image $s'$ of $s$ under $\sigma$ cannot
be orthogonal to $e_{1}'$ and thus intersects the plane $E'$ in
a point $x'$. This point $x'$ shall be the image of $x$ under $\varphi_{0}$.

The next three parts of the proof are dedicated to the investigation
of $\varphi_{0}$. After this has been done, $\varphi_{0}$ can be
continued from $E$ to a mapping of $H$ onto $H'$ satisfying assertion
(b) of the theorem.

\subsection*{Part 3}

The decisive point of the whole proof is a general construction that
will serve as the main tool for the investigation of $\varphi_{0}$
. Let $s$ be a ray of $H$ with a generating element\[
v=\sum_{I}a_{i}e_{i}\]
The equation of the ray plane $E(s)$ orthogonal to $s$ is\[
\sum_{I}a_{i}x_{i}=0\]
and the condition for an element\[
x=\sum_{I}x_{i}e_{i}\]
of $H$ to lie in the intersection\[
g(s)=E\cap E(s)\]
of $E$ and $E(s)$ is given by the equation

\begin{equation}
\sum_{J}a_{i}x_{i}+a_{1}=0\qquad\mbox{ with}\qquad J=I/\left\{ 1\right\} \end{equation}

Since $E(s)$ may be considered as a set of rays all of whom being
orthogonal to $s$, and, because $\sigma$ conserves orthogonality,
the image $E'(s')$ of $E(s)$ under $\sigma$ is the ray plane corresponding
to the image $s'$ of $s$. If\[
v'=\sum_{I}a_{i}'e_{i}'\]
is an element generating $s'$, then the equation for $E'(s')$ is

\[
\sum_{I}a_{i}'x_{i}'=0\]
Since $\varphi_{0}$ is based on $\sigma$ by definition and transforms
$E$ into $E'$, the set $g(s)$ is transformed into the set\[
g'(s')=E'\cap E'(s')\]
with the equation\begin{equation}
\sum_{J}a_{i}'x_{i}'+a_{1}'=0\end{equation}

\subsection*{Part 4}

The properties of the mapping $\varphi_{0}$ can be found by investigating
special cases. 

First of all let $v$ be the element\[
v=y_{i}e_{1}-e_{i}\]
Then $g(s)$ is given by the equation $x_{i}-y_{i}=0$ and thus consists
of all elements of $E$ with the same $i$.th coordinate $y_{i}$.
Since $v$ is orthogonal to all base elements of $H$ different from
$e_{1}$ and $e_{i}$ a generating element $v'$ of the ray $s'$
can be chosen as\[
v'=y_{i}'e_{1}'-e_{i}'\]
with the consequence that $g'(s')$ consists of all elements of $E'$
with the same $i$.th coordinate $y_{i}'$. Hence the whole mapping
$\varphi_{0}$ is decomposing into functions $f_{i}$ between the
coordinates, one for each index $i$. The coordinate $y_{i}$ is related
to the coordinate $y_{i}'$ by\[
y_{i}'=f_{i}(y_{i})\qquad i\epsilon J\]

and correspondingly\[
y_{j}'=f_{j}(y_{j})\qquad j\epsilon J\]

for another index $j$ different from 1 and $i$. By construction
of $\varphi_{0}$ two special values are\[
f_{i}(0)=0\qquad\mbox{ and}\qquad f_{j}(0)=0\]
\[
f_{i}(1)=c_{i}\qquad\mbox{ and}\qquad f_{j}(1)=c_{j}\]
Only for the purpose of this subsection new coordinates\[
x_{i}^{*}=c_{i}^{-1}\qquad\mbox{ and}\qquad x_{j}^{*}=c_{j}^{-1}\]
are introduced. Since all base elements different from $1$, $i$
and $j$ are irrelevant for the rest of this subsection, they can
be set to zero. Thus the problem of investigating $\varphi_{0}$ is
reduced to the analysis of a three dimensional vector space spanned
by $e_{1}$, $e_{i}$ and $e_{j}$.

Next let be \[
v=e_{i}-e_{j}\]
Then $g(s)$ is containing the line joining the points (1,0,0) and
(1,1,1) of $E$ and $g'(s')$ is containing the line joining the points
(1,0,0) and (1,1,1) of $E'$ given in the new coordinates. Hence the
two functions $f_{i}$ and $f_{j}$ are the same function $f$ and\begin{equation}
y_{i}^{*}=f(y_{i})\qquad i\epsilon J\end{equation}
\begin{equation}
y_{j}^{*}=f(y_{j})\qquad j\epsilon J\end{equation}
Next let be\[
v=\mu e_{1}+\lambda e_{i}-e_{j}\]
a generating vector of $s$. Then the equation for $g(s)$ can be
written as \begin{equation}
y_{j}=\lambda y_{i}+\mu\end{equation}
A generating vector of $s'$ is \[
v'=\mu'e_{1}'+\lambda'e_{i}'-e_{j}'\]
Hence the equation for $g'(s')$ can be written as

\begin{equation}
y_{j}^{*}=\lambda'y_{i}^{*}+\mu'\end{equation}

The combination of (3) and (4) with (5) and (6) will yield\[
f(\lambda y_{i}+\mu)=\lambda'f(y_{i})+\mu'\]

and especially\[
\mu'=f(\mu)\qquad\mbox{for}\qquad\ y_{i}=0\]
\[
\lambda'=f(\lambda)\quad\qquad\qquad\mbox{for}\qquad y_{i}=1,\,\mu=0\]
For arbitrary $a$ and $b$ one can conclude that\[
f(a+b)=f(a)+f(b)\qquad\mbox{with}\qquad a=\lambda,\, b=\mu,\, y_{i}=1\]
\[
f(ab)=f(a)f(b)\;\,\qquad\qquad\mbox{with}\qquad a=\lambda,\, b=y_{i},\,\mu=0\]
Thus $f$ is an automorphism on the field of complex numbers. 

Now one can go back to the original coordinates and write\[
y_{i}'=c_{i}f(y_{i})\qquad i\epsilon J\]
The complex number $c_{i}$ may be split into an absolute value $r_{i}$
and a phase factor $w_{i}$ according to\[
c_{i}=r_{i}w_{i}\qquad\qquad i\epsilon J\]
One can get rid of the phase factor $w_{i}$ by fixing the phase factor
$p_{i}$ in $e_{i}'$ that was introduced in part 2 of the proof such
that\[
y_{i}'=r_{i}f(y_{i})\qquad\qquad i\epsilon J\]
with $r_{i}$ being real and non negative.

\subsection*{Part 5}

Now the mapping $\varphi_{0}$ can be continued from $E$ to a mapping
$\varphi$ of the whole Hilbert space $H$. For this purpose three
cases should be distinguished.

\subsubsection*{Case 1}

If $x$ is lying on the coordinate axis $k_{1}$, then the coordinates
of the image $x'$ by definition shall be\[
x_{1}'=r_{1}f(x_{1})\qquad\mbox{ and}\qquad x_{i}'=0\;\mbox{ for}\;\; i\epsilon J\]
with $r_{1}=1$.

\subsubsection*{Case 2}

If $x$ is in the ray plane orthogonal to $k_{1}$ and in a ray $s$,
then the image $s'$ of $s$ is lying in the ray plane orthogonal
to $k_{1}'$. Moreover $x$ is contained in a subspace $T$ spanned
by $k_{1}$ and $s$. In $T$ there is an element $y$ of $E$ with
the same coordinate\[
y_{i}=x_{i}\qquad\qquad i\epsilon J\]

as $x$. The image $T'$ of $T$ contains the image $y'$ of $y$
and furthermore an element $x'$ with the same coordinates\[
x_{i}'=y_{i}'\qquad\qquad i\epsilon J\]

as $y'$. Then the element with the coordinates\[
x_{1}'=0\qquad\mbox{ and}\qquad x_{i}'=y_{i}'=r_{i}f(y_{i})=r_{i}f(x_{i})\;\mbox{ for}\;\; i\epsilon J\]
shall be the image of $x$ under $\varphi$.

\subsubsection*{Case 3}

If $x$ is neither in the plane with $x_{1}=0$ nor on the axis $k_{1}$,
then a ray $s$ containing $x$ intersects $E$ in an element $y$
with coordinates $y_{i}$ for $i\epsilon J$. The image $y'$ of $y$
with the coordinates\[
y_{1}'=1\qquad\mbox{ and}\qquad y_{i}'=r_{i}f(y_{i})\;\mbox{ for}\;\; i\epsilon J\]
is lying on the same ray $s'$ as the element $x'$ of $H'$ with
the coordinates\[
x_{1}'=f(x_{1})\qquad\mbox{ and}\qquad x_{i}'=f(x_{1})r_{i}f(y_{i})=r_{i}f(x_{1}y_{i})=r_{i}f(x_{i})\;\mbox{ for}\;\; i\epsilon J\]
This element $x'$ shall be the image of $x$ under $\varphi$.

In all three cases one can write\[
x_{i}'=r_{i}f(x_{i})\qquad\qquad i\epsilon I\]
The stepwise construction of the mapping $\varphi$ was merely based
on the ray mapping $\sigma$. Hence the result reproduces $\sigma$
in the sense that $\varphi(x)\epsilon\sigma(x)$, if $x\epsilon s$.

\subsection*{Part 6}

Until now the invariance of orthogonality under the ray mapping $\sigma$
was sufficient. But now the conservation of the ray function $u$
is needed.

For an element\[
h_{i}=e_{1}+x_{i}e_{i}\qquad\qquad i\epsilon J\]
of $H$ the value $|x_{i}|^{2}/(1+|x_{i}|^{2})$ is equal to the value
$u(k_{i},s)$ of the ray function applied to the coordinate axis $k_{i}$
and the ray $s$ containing the element $h_{i}$. Correspondingly
the absolute value $|x_{i}'|^{2}/(1+|x_{i}'|^{2})$ is equal to $u(k_{i}',s')$
for the images $k_{i}'$ and $s'$ of $k_{i}$ and $s$. Conservation
of the ray function will yield\begin{equation}
|x_{i}'|=|x_{i}|\qquad\qquad i\epsilon J\end{equation}
or\[
|r_{i}f(x_{i})|=|x_{i}|\qquad\qquad i\epsilon J\]

and especially\[
|r_{i}f(1)|=|r_{i}|=1\qquad\qquad i\epsilon J\]

Thus\[
|f(x_{i})|=|x_{i}|\]

That is to say, the automorphism $f$ leaves invariant the absolute
values of all coordinates $x_{i}$ and hence all non-negative real
numbers. Additionally one has\[
f(-1)+1=f(-1)+f(1)=f(0)=0\]
and thus\[
f(-1)=-1\]
As a conclusion $f$ is the identity on the field of real numbers.
Because $i$ and $-i$ are the only solutions of the equation\[
x^{2}+1=0\]
they only underly a permutation. Thus $f$ is either the identy or
the transition to the complex conjugate.

\section{Comparison with other proofs}

Part 1 of the proof given in section 4 coincides with the corresponding
part in the proof of S. Weinberg {[}2]. But in contrast to the strategy
that was pursued here the proof of Weinberg determines the phase factors
already at an early stage. Thus Weinberg has to fight with a lot of
problems as for instance with the discrimination and investigation
of several cases. If the text written down in {[}2] is taken together
with all footnotes and all calculations, whose explications are lacking,
then the proof is rather complicated.

The proof of K. Keller {[}3] has in common with the ansatz given here
that it associates an investigation concerning Hilbert spaces with
analytic geometry and not, as usual, with functional analysis. But
the combination with projective geometry in arbitrary dimensions seems
to be a detour, especially because the main theorem of projective
geometry can be reduced to the corresponding theorem of affine geometry.
In the present paper it has been shown that simple geometry in two
or three dimensions is sufficient for a proof of Wigner' s theorem.

\section{Summary}

The leading idea of the proof given here was to reduce the general
case to some simple geometry in two or three dimensions. For this
purpose a general construction was implemented using as tool some
simple concepts and the assumption of orthogonality. The geometry
in three dimensions is thus on the one hand an essential part of the
proof and on the other hand a special example well suited to illustrate
the typical features of Wigners theorem.

\section{Final remark}

I wish to thank Prof. Dr. Fredenhagen and the working group 'Algebraic
Quantum Field Theory' at the II. Institute for Theoretical Physics
of Hamburg University for the opportunity to present the content of
this paper and for discussions. Furthermore I am indebted to Hinnerk
Albert for support in technical detail.

\end{document}